\documentclass[epj,nopacs]{svjour}
\usepackage{amsmath}
\usepackage{graphicx}
\usepackage{subfig} % subfig.sty for a double column floating figure using two subfigures
\usepackage{url}

\def\aprle{\buildrel < \over {_{\sim}}}
\def\aprge{\buildrel > \over {_{\sim}}}
\newcommand{\ANKARA}      {28}
\newcommand{\ANNECY}      {9}
\newcommand{\BARI}        {33}
\newcommand{\BARIINFN}    {20}
\newcommand{\BERN}        {4}
\newcommand{\BOLOGNA}     {13}
\newcommand{\BOLOGNAINFN} {14}
\newcommand{\BRUSSELS}    {36}
\newcommand{\DUBNA}       {22}
\newcommand{\FRASCATI}    {16}
\newcommand{\FUNABASHI}   {26}	
\newcommand{\GAZWADONG}   {31}
\newcommand{\HAIFA}       {27}
\newcommand{\HAMBURG}     {10}
\newcommand{\KARIYA}      {32}
\newcommand{\KOBE}        {3}
\newcommand{\LAQUILA}     {21}
\newcommand{\LNGS}        {18}
\newcommand{\LYON}        {5}
\newcommand{\MOSCOWINR}   {1}
\newcommand{\MOSCOWITEP}  {23}
\newcommand{\MOSCOWLPI}   {7}
\newcommand{\MOSCOWSINP}  {2}
\newcommand{\MUNSTER}     {24}
\newcommand{\NAGOYA}      {25}
\newcommand{\NAPOLI}      {19}
\newcommand{\NAPOLIINFN}  {15}
\newcommand{\PADOVA}      {12}
\newcommand{\PADOVAINFN}  {8}
\newcommand{\ROMA}        {34}
\newcommand{\ROSTOCK}     {29}
\newcommand{\SALERNO}     {11}
\newcommand{\STRASBOURG}  {17}
\newcommand{\UTSUNOMIYA}  {35}
\newcommand{\ZAGREB}      {30}
\newcommand{\ZURICH}      {6}
\newcommand{\OperaInstitutes}{
\MOSCOWINR   . INR-Institute for Nuclear Research of the Russian Academy of Sciences, RUS-117312 Moscow, Russia\\
\MOSCOWSINP  . SINP MSU-Skobeltsyn Institute of Nuclear Physics of Moscow State University, RUS-119992 Moscow, Russia \\
\KOBE        . Kobe University, J-657-8501 Kobe, Japan \\
\BERN        . Albert Einstein Center for Fundamental Physics, Laboratory for High Energy Physics (LHEP), University of Bern, CH-3012 Bern, Switzerland \\
\LYON        . IPNL, Universit\'e Claude Bernard Lyon 1, CNRS/IN2P3, F-69622 Villeurbanne, France\\
\ZURICH      . ETH Zurich, Institute for Particle Physics, CH-8093 Zurich, Switzerland \\
\MOSCOWLPI   . LPI-Lebedev Physical Institute of the Russian Academy of Sciences, RUS-117924 Moscow, Russia\\
\PADOVAINFN  . INFN Sezione di Padova, I-35131 Padova, Italy \\
\ANNECY      . LAPP, Universit\'e de Savoie, CNRS/IN2P3, F-74941 Annecy-le-Vieux, France\\
\HAMBURG     . Hamburg University, D-22761 Hamburg, Germany\\
\SALERNO     . Dipartimento di Fisica dell'Universit\`a  di Salerno and INFN, I-84084 Fisciano, Salerno, Italy \\
\PADOVA      . Dipartimento di Fisica dell'Universit\`a  di Padova, I-35131 Padova, Italy \\
\BOLOGNA     . Dipartimento di Fisica dell'Universit\`a  di Bologna, I-40127 Bologna, Italy \\
\BOLOGNAINFN . INFN Sezione di Bologna, I-40127 Bologna, Italy \\
\NAPOLIINFN  . INFN Sezione di Napoli, 80125 Napoli, Italy \\
\FRASCATI    . INFN - Laboratori Nazionali di Frascati dell'INFN, I-00044 Frascati (Roma), Italy \\
\STRASBOURG  . IPHC, Universit\'e de Strasbourg, CNRS/IN2P3, F-67037 Strasbourg, France\\
\LNGS        . INFN - Laboratori Nazionali del Gran Sasso, I-67010 Assergi (L'Aquila), Italy \\
\NAPOLI      . Dipartimento di Fisica dell'Universit\`a Federico II di Napoli, 80125 Napoli, Italy \\
\BARIINFN    . INFN Sezione di Bari, I-70126 Bari, Italy \\
\LAQUILA     . Dipartimento di Fisica dell'Universit\`a dell'Aquila and INFN, I-67100 L'Aquila, Italy \\
\DUBNA       . JINR-Joint Institute for Nuclear Research, RUS-141980 Dubna, Russia \\
\MOSCOWITEP  . ITEP-Institute for Theoretical and Experimental Physics, RUS-117218 Moscow, Russia \\
\MUNSTER     . University of M\"unster, D-48149 M\"unster, Germany\\
\NAGOYA      . Nagoya University, J-464-8602 Nagoya, Japan\\
\FUNABASHI   . Toho University, J-274-8510 Funabashi, Japan \\
\HAIFA       . Department of Physics, Technion, IL-32000 Haifa, Israel\\
\ANKARA      . METU-Middle East Technical University, TR-06531 Ankara, Turkey \\
\ROSTOCK     . Fachbereich Physik der Universit\"at Rostock, D-18051 Rostock, Germany\\
\ZAGREB      . IRB-Rudjer Boskovic Institute, HR-10002 Zagreb, Croatia\\
\GAZWADONG   . Gyeongsang National University, 900 Gazwa-dong, Jinju 660-701, Korea\\
\KARIYA      . Aichi University of Education, J-448-8542 Kariya (Aichi-Ken), Japan\\
\BARI        . Dipartimento di Fisica dell'Universit\`a  di Bari, I-70126 Bari, Italy \\
\ROMA        . Dipartimento di Fisica dell'Universit\`a  di Roma ``La Sapienza'' and INFN, I-00185 Roma, Italy \\
\UTSUNOMIYA  . Utsunomiya University, J-321-8505 Tochigi-Ken, Utsunomiya, Japan\\
\BRUSSELS    . IIHE, Universit\'e Libre de Bruxelles, B-1050 Brussels, Belgium
}
\newcommand{\OperaAuthorList}{
N.~Agafonova\inst{\MOSCOWINR},
A.~Anokhina\inst{\MOSCOWSINP},
S.~Aoki\inst{\KOBE},
A.~Ariga\inst{\BERN},
T.~Ariga\inst{\BERN},
D.~Autiero\inst{\LYON},
A.~Badertscher\inst{\ZURICH},
A.~Bagulya\inst{\MOSCOWLPI},
A.~Bertolin\inst{\PADOVAINFN},
M.~Besnier\inst{\ANNECY,}\thanks{Now at Laboratoire Leprince-Ringuet - \'Ecole polytechnique, 91128 Palaiseau Cedex (France)}
% ...
}

\begin{document}

\hugehead

\title{Measurement of the atmospheric muon charge ratio with the OPERA detector}

\author{\OperaAuthorList}
\institute{\OperaInstitutes}

%\date{Received: date / Revised version: date}
\date{}

\abstract{
The OPERA detector at the Gran Sasso underground laboratory (LNGS) was used to measure the atmospheric muon charge ratio $R_\mu = N_{\mu^{+}}/N_{\mu^{-}}$ in the TeV energy region. We analyzed 403069 atmospheric muons corresponding to 113.4 days of livetime during the 2008 CNGS run. We computed separately the muon charge ratio for single and for multiple muon events in order to select different energy regions of the primary cosmic ray spectrum and to test the $R_\mu$ dependence on the primary composition. The measured $R_\mu$ values were corrected taking into account the charge-misidentification errors.
Data have also been grouped in five bins of the ``vertical surface energy'' $\mathcal{E}_\mu \cos \theta$. A fit to a simplified model of muon production in the atmosphere allowed the determination of the pion and kaon charge ratios weighted by the cosmic ray energy spectrum.
}

%\titlerunning{Measurement of the cosmic ray muon charge ratio with the OPERA detector}
%\authorrunning{N.~Agafonova \emph{et al.}}

\maketitle

\section{Introduction}
\label{sec:intro}

Primary cosmic rays (typically protons) impinging on the Earth's atmosphere
produce showers of secondary particles which propagate down to the ground
level. Most of the interaction products are $\pi$ and $K$ mesons which in turn
decay or interact, depending on their energy and on the air density profile they
pass through. The decay of $\pi^{0}$ mesons gives rise to the electromagnetic component of
the showers, the decay of $\pi^{\pm}$ and $K^{\pm}$ mesons yields mostly muons which are the most 
penetrating charged particles and therefore the most abundant charged component at sea level.
In particular only the most energetic muons can penetrate deep underground. 
The Gran Sasso laboratory (LNGS) is located at an average depth of 3800 m.w.e.
and the minimum muon energy required to reach the underground depth is around 1.5 TeV 
while the residual underground energy is about 270 GeV averaged over all directions \cite{trd}.

The muon charge ratio $R_{\mu} = N_{\mu^{+}}/N_{\mu^{-}}$ defined as the number of positive over negative charged muons,
results from several contributions: the primary cosmic ray
composition (in particular the ratio of protons over heavier primaries), 
hadronic-interaction features, atmospheric conditions (negligible above a few GeV)
and, at very high energy, the contribution of muons from charmed particle decays (prompt muons) \cite{bugaev}.
The muon charge ratio at sea level was extensively studied in the past since it is an indicator of 
important aspects of cosmic rays and particle physics.

An exhaustive compilation of measurements in a wide energy range is reported in Ref. \cite{grieder}. 
In the interval from a few hundred MeV to 300 GeV the muon charge ratio $R_{\mu}$ stays around 1.27. At higher energies several competing processes can affect its value. 
Since strong interaction production channels lead to a $K^+/K^-$ ratio higher than the $\pi^+/\pi^-$ one and the fraction of muons from kaon decays increases with the energy,
the muon charge ratio is expected to rise as the energy increases. On the other hand, as the zenith angle increases and hence longer lived mesons have a higher probability to decay in the less dense layers of the high atmosphere the fraction of muons from pion decay increases and the muon charge ratio decreases.
We also expect a dependence of the muon charge ratio on the underground muon multiplicity $n_\mu$, which is related to the energy of the primary cosmic rays and to their chemical composition. 
For primaries different from protons the positive charge excess is reduced and so is the muon charge ratio \cite{icrc09-muraro}.

A simplified model of the atmospheric muon charge ratio is obtained from the muon spectrum \cite{gaisser}
\begin{equation}
\Phi_{\mu} = \frac{\Phi_N(\mathcal{E}_{\mu})}{1-Z_{NN}} \sum_{i=1}^{N_{par}} \frac{a_i Z_{Ni}}{1+b_i \mathcal{E}_\mu /\epsilon_i(\theta)}
\label{eq:0}
\end{equation}
where $\Phi_N(\mathcal{E}_{\mu}) \simeq \Phi_0 \mathcal{E}_{\mu}^{-(\gamma + 1)}$ is the primary spectrum of nucleons (evaluated at the muon 
energy in the atmosphere $\mathcal{E}_{\mu}$) with a spectral index $\gamma + 1 \simeq 2.7$.
For each of the $N_{par}$ muon parents ($\pi^{}$, $K$, charmed particles etc.) the constants $a_i$ and $b_i$ contain the kinematical factors for the decay into muons, $\epsilon_i(\theta)$ are the critical energies defined as the energies above which interaction processes dominate over decay. They depend on the ratio $m_i/\tau_i$ (mass over rest lifetime of the muon parent) and on the atmospheric profile density and therefore on the zenith angle $\theta$. A good approximation for $\epsilon_i(\theta)$ which takes into account the Earth's curvature is
\begin{equation}
\epsilon_i(\theta) = \frac{\epsilon_i(0)}{\cos \theta^{*}}
\label{eq:ecrit}
\end{equation}
with
\begin{equation}
\cos \theta^{*} = \sqrt{1-\sin^2\theta \left(\frac{R_e}{R_e+h}\right)^2}
\label{eq:costar}
\end{equation}
where $R_e$ is the Earth's radius and $h$ is the muon production height. By using eq. \ref{eq:costar} the zenith angle is evaluated at the muon production point and not at the detector site. By choosing $h = 30$ km an agreement within 5\% with the precise $\epsilon_i(\theta)$ computation is obtained \cite{lipari1}.

The spectrum weighted moments are defined as:
\begin{equation}
Z_{ij} = \int^{1}_{0} \frac{1}{\sigma_{ij}}\frac{d\sigma_{ij}}{dx_{lab}}(x_{lab})^{\gamma-1}dx_{lab}
\label{eq:3}
\end{equation}
where $\sigma_{ij}$ is the inclusive cross-section for the production of a particle $j$ from the collision of a particle $i$ with a nucleus in the atmosphere and $x_{lab} = E_j/E_i$ is the energy fraction carried by the secondary particle. 
Use of the $Z$ factors shows explicitly that particle production in cosmic ray cascades is concentrated in the forward fragmentation region at large $x_{lab}$.

From Eq. \ref{eq:0} the $i$-th contribution to the muon flux is proportional to $Z_{Ni}$ and is suppressed for $\mathcal{E}_\mu \gg \epsilon_i(\theta)$.
Therefore each contribution to the muon charge ratio produced by different muon parents can be disentangled by studying
the muon charge ratio as a function of the muon energy $\mathcal{E}_\mu$.

Considering only $\pi$ and $K$ parent mesons Eq. \ref{eq:0} can be separated for $\mu^{+}$ and $\mu^{-}$
\begin{equation}
\Phi_{\mu^{\pm}} \propto \biggl( \frac{a_\pi Z_{N\pi^{\pm}}}{1+b_\pi \mathcal{E}_\mu \cos \theta^{*} /\epsilon_{\pi}} + 
\frac{a_K Z_{NK^{\pm}}}{1+b_K \mathcal{E}_\mu \cos \theta^{*} /\epsilon_{K}} \biggr)
\label{eq:2}	
\end{equation}
where $\epsilon_{\pi} \simeq 115$ GeV and $\epsilon_{K} \simeq 850$ GeV are the pion and kaon critical energies along the vertical direction, respectively.
These energies have to be compared with the corresponding value for charmed particles, $\epsilon_{X} > 10^7$ GeV. The prompt muon component (from charmed particles) is therefore isotropically distributed since
the corresponding $\cos \theta^{*}$ factor is suppressed, at least in the TeV region.

Eq. \ref{eq:2} contains most of the aspects already discussed. First we note that the correct variable to describe the
evolution of the charge ratio is the product $\mathcal{E}_\mu \cos \theta^{*}$, the ``vertical surface energy'' \cite{icrc69-parker,icrc07-goodman}. 
Moreover the two energy scales which determine the pion and kaon contributions to $R_\mu$ are the critical energies $\epsilon_{\pi}$ and $\epsilon_K$. 
The evaluation of the muon surface energy $\mathcal{E}_\mu$ depends on the rock depth crossed by the muon to reach the detector and therefore
the distribution of $\mathcal{E}_\mu \cos \theta^{*}$ is related to the shape of the overburden.
Measurements of the muon charge ratio at high energies and large zenith angles, corresponding to $\langle \mathcal{E}_\mu \cos \theta^{*} \rangle \sim$ 0.5 TeV, are given in Ref. \cite{utah}. More recent data with large statistics at $\langle \mathcal{E}_\mu \cos \theta^{*} \rangle \sim$ 1 TeV are presented in Ref. \cite{minos}. These results suggest a smooth transition toward the energy region where kaon contribution becomes significant.

The LNGS laboratory is located at $\langle \mathcal{E}_\mu \cos \theta^{*} \rangle \simeq$ 2 TeV, well above the kaon critical energy $\epsilon_K$.
This allows the measurement of the ratio $Z_{NK^+}/Z_{NK^-}$ whose value is poorly known in the fragmentation region. This has also a strong impact, for instance on the evaluation of the flux of TeV atmospheric neutrinos, which are dominated by kaon production.
OPERA is the first large magnetized detector that can measure the muon charge ratio at the LNGS depth, with an acceptance for cosmic ray muons coming from above $\mathcal{A}$ = 599 m$^2 \cdot$ sr ($\mathcal{A}$ = 197 m$^2$ sr for muons crossing the spectrometer sections).

The paper is organized as follows. In Sec. \ref{sec:opera} the detector is briefly described, while in Sec. \ref{sec:analysis} we describe data selection and reconstruction, Monte Carlo simulation and data reduction. In Sec. \ref{sec:cr} the muon charge ratio at the LNGS underground depth and its systematic error are evaluated. Finally in Sec. \ref{sec:final} we give $R_\mu$ as a function of the underground momentum and of the variable $\mathcal{E}_\mu \cos \theta^{*}$ fitted to Eq. \ref{eq:2}.

\begin{figure*}[t]
\begin{center}
\includegraphics[width=1.75\columnwidth]{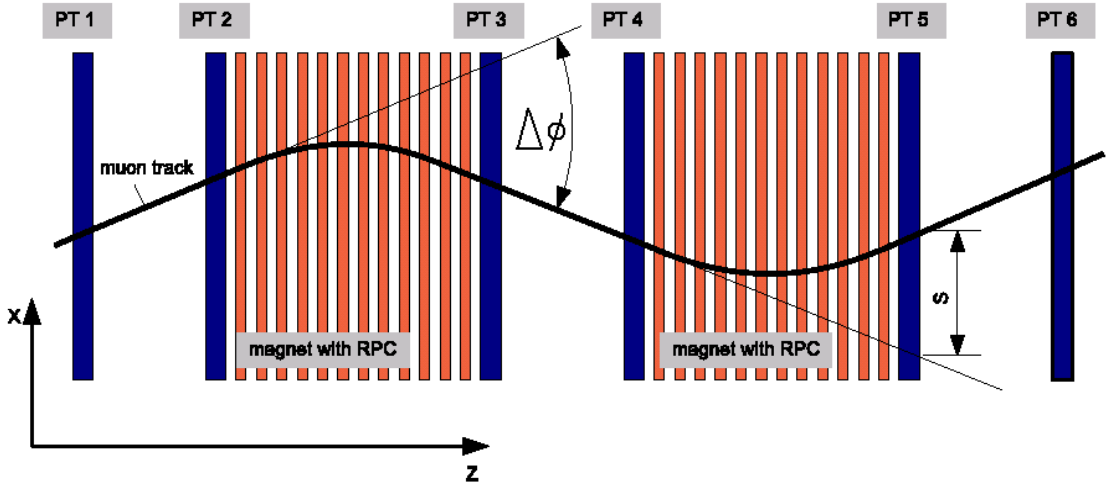}
\end{center}
\caption{Schematic view of a charged particle crossing one spectrometer. The six PT stations are shown in dark grey; the 24 iron slabs (12 per arm) interleaved with 22 RPC planes are shown in light grey. Each spectrometer arm provides an independent measurement of charge/momentum, provided the track is reconstructed in at least one station (or station doublet) in each side of the arm.}
\label{fig:magnet}
\end{figure*}

\section{The OPERA detector}
\label{sec:opera}

OPERA is a hybrid experiment with electronic detectors and nuclear emulsions located in Hall C of the underground Gran Sasso Laboratory in central Italy \cite{operaDet}.
The main physics goal of the experiment is to observe neutrino flavor oscillations through the appearance of $\nu_\tau$ neutrinos in the $\nu_\mu$ CNGS beam \cite{cngs}.
The detector design was optimized to identify the $\tau$ lepton via the topological observation of its decay: this requires a target mass of more than a kton to maximize the neutrino interaction probability and a micrometric resolution to detect the $\tau$ decay. To accomplish these requirements the detector concept is based on the Emulsion Cloud Chamber (ECC) technique combined with real-time detection techniques (``electronic detectors'').

The ECC basic unit in OPERA is a ``brick'' made of 56 lead plates, 1 mm thick, providing the necessary mass to cope with the small neutrino cross-section, interleaved with 57 nuclear emulsion films (industrially produced), providing the necessary spatial and angular resolution to identify tau decay topologies. In total, 150000 bricks have been assembled reaching the overall 
mass of 1.25 kton.

The electronic detectors are used to trigger the neutrino interactions, to locate the brick in which the interaction occurred, to identify muons and measure their charge and  momentum.

The detector is composed of two identical parts called supermodules (SM), each one consisting of a target section and a magnetic spectrometer.
In the target the bricks are arranged in 29 vertical ``walls'' transverse to the beam direction interleaved with electronic Target Tracker (TT) walls. Each TT wall consists of a double layer of 256 scintillator strips, for vertical and horizontal coordinate measurements.
The TTs can trigger the data acquisition and locate the brick in which the interaction occurred.

The target section is followed by a magnetic spectrometer (see Fig. \ref{fig:magnet}). A large dipolar iron magnet is instrumented with Resistive Plate Chambers (RPCs). The magnetic field is 1.53 T, directed vertically transverse to the neutrino beam axis. The RPC planes are inserted between the iron slabs: they provide the tracking inside the magnet and the range measurement for stopping muons \cite{rpc}.

The deflection of charged particles in the magnet is measured by six stations of vertical drift tubes, the Precision Trackers (PT),
grouped in 3 pairs placed upstream of the first, in between and downstream of the second magnet arms (Fig. \ref{fig:magnet}).
Each PT station is formed by four staggered layers of aluminum tubes, 8 m long, with 38 mm outer diameter. The spatial resolution is better than 300 $\mu$m in the bending (horizontal) plane: this allows the determination of the muon charge sign with high accuracy, and the momentum measurement with a resolution of better than 20\% for momenta $<\,$50 GeV/c for charged particles coming from the CNGS direction \cite{hpt}. The PT system is triggered by the RPC timing boards with a configuration optimised to collect both beam and cosmic ray muons with high efficiency.

In order to remove ambiguities in the reconstruction of particle trajectories, in particular in multi-track events, each spectrometer is instrumented with additional RPC planes (XPC), with two crossed strip planes rotated by $\pm 42.6^\circ$ with respect to the horizontal.

Two RPC planes (VETO) are placed in front of the detector, acting as a veto for charged particles originating from the upstream material (mainly muons from neutrino interactions in the rock).

Finally we emphasize that for this analysis the OPERA detector was used differently from what it was conceived for. This is particularly true for the PT system which was configured and optimized to reconstruct and measure particles traveling along the CNGS direction.

\section{Data analysis}
\label{sec:analysis}

\subsection{Data collection and selection}
\label{subsec:collection}

The results presented are based on data recorded during the CNGS physics run, from June 18th until November 10th 2008. 
The detector ran in the standard configuration, with the magnetic field directed along the vertical axis in the first arm of both spectrometers, and in the opposite direction in the second arm. Moreover a sample of cosmic ray muons was collected with the magnetic field switched off in order to improve the alignment between PT stations and to evaluate systematic uncertainties. A limited data sample was obtained inverting the magnet polarity to cross check the charge reconstruction.

The data acquisition was segmented in ``extraction periods'' of about 12 hours each. 
Only periods of data taking where all the main 	detector subsystems ran in stable 
conditions were considered. They amount to about 78\% of the total duration of the run.
The total number of selected events is 403069 corresponding to 113.4 days of livetime.

\subsection{Event reconstruction}
\label{subsec:reco}
The OPERA standard software for beam event reconstruction was complemented with a set of dedicated software tools developed for cosmic ray events. Once the event is tagged as ``off-beam'', that is outside the CNGS spill window it is classified as {\it cosmic} and processed in a dedicated way. This choice was required by the different topologies of beam and cosmic ray events. Beam events come from a well defined direction (the CNGS one) and the reconstruction code is optimized to follow a single long track (the muon escaping from the neutrino-interaction region) on the $z$-axis. Cosmic ray events come from all directions, they are not generated within the target and a fraction of them ($\sim$ 5\% in OPERA) are muon bundles. A brief description of the code follows.

The reference frame is defined to have the $z$ axis along the Hall C longitudinal direction (from north to south), $y$ perpendicular to the floor pointing toward the zenith and $x$ describing a right-handed frame. In this coordinate system, the zenith direction $\theta$ is defined by the angle with the $y$ axis, the azimuth direction $\phi$ by the angle with the $z$ axis. Event reconstruction is performed separately in the two projected views $T_{xz}$ and $T_{yz}$.
First the event direction is determined by using the Hough transform. Using a Monte Carlo simulation we estimated an angular resolution of better than 0.5$^{\circ}$, both in the $\theta$ and $\phi$ directions, for single as well as for multiple track events. Then, the direction information is used to subdivide the $T_{xz}$ and $T_{yz}$ views in slices 25 cm wide having the same slant as the reconstructed direction. The hits within the same slice are then processed separately to search for a track ``seed'' of at least three aligned points. If a seed is found, all the other hits in the corresponding projected view are linked to the selected track according to pre-defined tolerances. Tracks independently reconstructed in each view are then merged together to build the three-dimensional event.

For this analysis, particular attention was devoted to the reconstruction of tracks with the Precision Tracker. A description of the PT system is available in \cite{hpt}, while the reconstruction procedures are detailed in \cite{bjoern}. Here we briefly mention the main steps used to extract charge and momentum from PT hits.

A muon crossing the spectrometer is deflected in the horizontal plane (Fig. \ref{fig:magnet}). Let $\phi$ be the angle between the particle direction and the $z$ axis, then the deflection $\Delta \phi$ is the difference between the two angles measured at both sides of each magnet arm.
Each $\phi$ value is obtained by fitting the spatial information provided by the PT system which is made of 8 layers of drift tubes arranged in two ``stations''.
Since cosmic ray tracks are almost uniformly distributed in $\phi$, many of them traverse only one single station\footnote{A PT station is made of 4 layers of drift tubes. A track is reconstructed when $N_{tubes} \ge$ 4.} of a pair and we refer to them as {\it singlets}, otherwise we call them {\it doublets}.

Due to the different lever arm the angular resolution for singlets is worse than for doublets.
For tracks parallel to the $z$-axis ($\phi = 0$) it is $\sigma_{\phi} \simeq 1$ mrad for singlets and $\simeq 0.15$ mrad for doublets.
Since for tilted tracks the number of fired tubes and their mutual distances are larger than for tracks with $\phi = 0$, the errors on the slopes decrease.
In order to increase the statistics we decided to consider both singlets and doublets: the percentage of cases in which both angles are reconstructed from doublets is $\sim$55\% of the total, $\sim$9\% are from singlets and the remaining 36\% are from mixed configurations, namely cases where an angle is reconstructed from a doublet and the other angle from a singlet.

Considering tracks with $\phi = 0$ and the Multiple Coulomb Scattering (MCS) within one magnet arm the total uncertainty on $\Delta \phi$ is
\begin{equation}
\sigma_{\Delta \phi} = \sqrt{\sigma^{2}_{\phi_1} + \sigma^{2}_{\phi_2} + \left( \frac{0.0136}{p} \right) ^{2}\frac{d}{X_0}}
\end{equation}
where $d$ = 0.6 m is the magnetized iron thickness, $X_0$ = 0.0176 m is the iron radiation length and $p$ is expressed in GeV/c.
Since the deflection due to the magnetic field is
\begin{equation}
\Delta \phi_B = \frac{0.3 B d}{p}
\end{equation}
where $B$ = 1.53 T, the requirement $\Delta \phi / \sigma_{\Delta \phi} > 1$ provides an estimate for the maximum detectable momentum, $p_{max} \simeq$ 1.25 TeV for doublets, $p_{max} \simeq$ 190 GeV for singlets and $p_{max} \simeq$ 260 GeV for mixed configurations.

For muon momenta $p \ll p_{max}$ the measurement error can be neglected and the only contributions to the $\Delta \phi$ uncertainty come from the MCS. 
In this ideal case the ratio $\Delta \phi_B/\Delta \phi_{MCS} \sim$ 3.5 corresponds to a charge-misidentification $\eta$ (defined as the fraction of tracks reconstructed with wrong charge sign) below 10$^{-3}$. In reality there are other effects which spoil the resolution and therefore the charge identification capability, as detailed in Sec. \ref{subsec:cr}.

To measure the charge and the momentum of a particle at least one $\Delta \phi$ angle is needed (for tracks parallel to the $z$-axis there can be up to 4 independent angles).
For each reconstructed $\Delta \phi_i$ the track momentum is computed by using the formula \cite{bjoern}
\begin{equation}
p_i = \frac{l(dE/dz)}{1-\exp[\Delta \phi_i(dE/dz)/e\bar{B}]} \sqrt{1+\frac{s_{yz}^2}{1+s_{xz}^2}}
\label{eq:momentum}
\end{equation}
where $l$ = 0.82 m is the total arm length (including RPC gaps), $\bar{B} = Bd/l$ is the effective magnetic field and $dE/dz$ is the ionization energy loss in the magnetized iron (which depends logarithmically on the muon momentum). The term in the square root takes into account the track slope $s_{yz}$ in the $T_{yz}$ view. The muon charge is determined from the sign of the $\Delta \phi_i$ angle, accounting for the particle arrival direction and the field orientation in the arm. The final muon momentum and charge are computed as the weighted average of the independent measurements.

\begin{figure*}[t]
   \centerline{\subfloat[Functional form of $M(\phi)$]{\includegraphics[width=1.0\columnwidth]{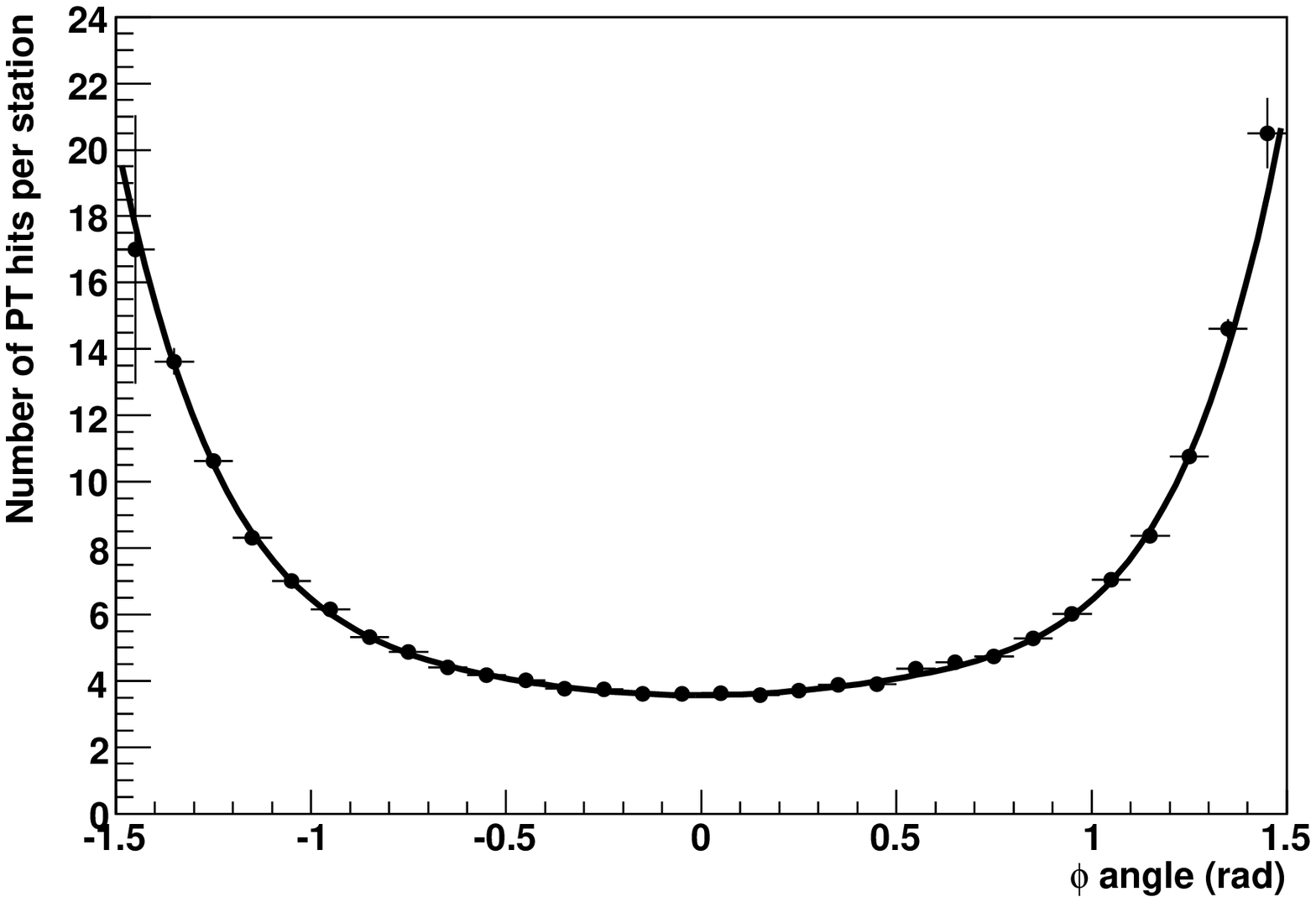} \label{fig:nDTcut1}}
              \hfil
              \subfloat[Rescaled data]{\includegraphics[width=1.0\columnwidth]{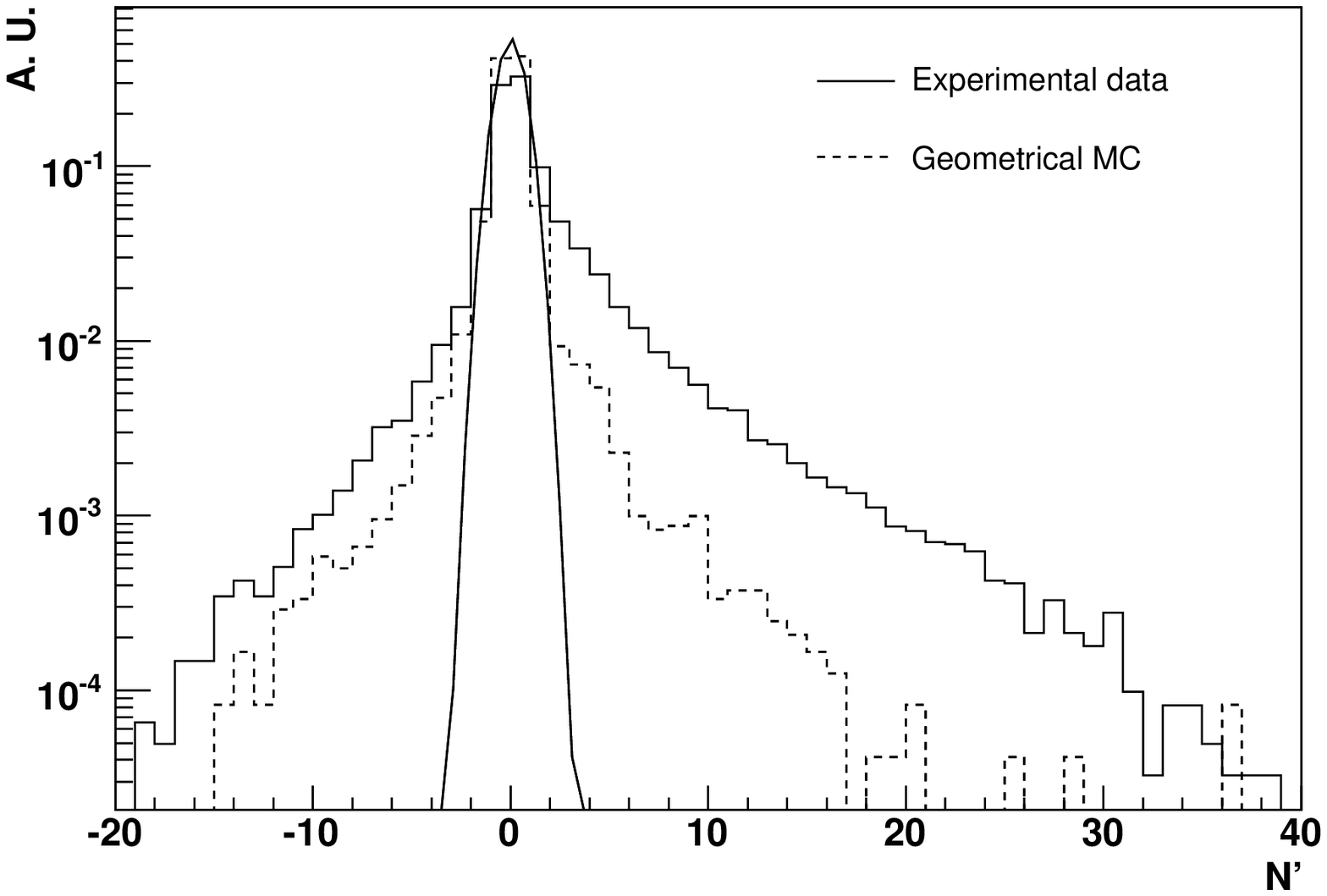} \label{fig:nDTcut2}}
             }
   \caption{Cut on the number of fired PT tubes/station. The geometrical dependence of the number of fired tubes on the $\phi$ angle is shown in (a), together with a polynomial fit; the rescaled distributions are shown in (b). A 3$\sigma$ cut of the Gaussian fit to Monte Carlo events where secondary particle production was switched off, was applied to the rescaled data (see text).}
   \label{fig:nDTcut}
\end{figure*}

\subsection{Monte Carlo simulation}
\label{subsec:mc}
Two different Monte Carlo simulations have been used in order to meet two different and conflicting requirements. From one side, one needs a large statistical sample of underground cosmic ray muons for calibration purposes and to correct for detector effects. This task was accomplished with a code which generates muons directly at the detector level fast enough to produce the required statistical amount of events. On the other side, information on cosmic ray primaries and the links between underground and surface variables required the use of a complete generator which simulates the full cosmic ray cascade in the atmosphere.

The first Monte Carlo code (MC1 hereafter) is based on a fast parameterized event generator developed in the framework of the MACRO experiment \cite{macro} and adapted for OPERA.
The generator considers the MACRO primary cosmic ray composition model \cite{macroFit} and, for each primary mass, it extracts from built-in probability tables the underground muon multiplicity. This choice ensures a self-consistency on the predicted muon flux underground since the composition model and the probability tables were obtained using the same hadronic interaction model. This means that the systematic errors on the primary composition and on the interaction model cancel and the Monte Carlo generator predicts the correct muon flux in the Hall B of LNGS.
The event direction ($\theta,\phi$) is sampled, together with the muon radial distribution with respect to the shower axis and the muon underground momentum. Finally, the event
kinematics is processed within the OPERA standard software chain, from the generation of the relevant physical processes in the electronic detectors up to the reconstruction level. 
The atmospheric muon charge ratio is introduced by hand ($R_\mu$ = 1.4).
Taking into account the livetime normalization, the ratio between OPERA data and Monte Carlo MC1 predictions is  
\begin{equation}
\frac{\textrm{Rate}_{REAL}}{\textrm{Rate}_{MC1}} = (95.9 \pm 0.3)\%
\label{eq:ratecfr}
\end{equation}
The difference from unity of the ratio is mainly due to the subdetectors efficiency mismatch between data and detector simulation.

The second Monte Carlo program (MC2) is based on the more detailed and physics-inspired Monte Carlo program described in Ref. \cite{blois08-sioli}, based on the FLUKA code \cite{fluka}. Here all the main physical processes are implemented, from the primary cosmic ray interactions and shower propagation in the atmosphere up to the muon transport in the overburden. The primary composition model is described in Ref. \cite{hoerandel} based on a global fit of several experimental observations.
This code is predictive of the muon charge ratio and allows the study of the relations between underground and surface muon energies. Due to its complexity the event production with MC2 is statistically limited.

In the following with the term ``Monte Carlo''  we refer to MC1 unless otherwise specified.

\subsection{Data reduction}
\label{subsec:reduction}

A set of data quality cuts were made in order to isolate a clean sample of reconstructed muon events. First at least one reconstructed $\Delta \phi$ angle is required for each event ({\it acceptance cut}). Then events with a large number of PT hits potentially dangerous for the muon charge determination are removed ({\it clean PT cut}). This typically occurs when some drift tubes are fired by secondary particles ($\delta$-rays, showers etc.) and the best $\chi^2$ track could result from a fake tube configuration.
In order to evaluate the maximum number of fired tubes/track allowed by geometrical considerations a special version of MC1 switching off delta ray and secondary particle production was run. By naming $M$ and $N$ the number of fired tubes from Monte Carlo simulation and experimental data, respectively, we derived the functional form $M = M(\phi)$, a six-order polynomial shown in Fig. \ref{fig:nDTcut1}.
$M(\phi)$ was used to rescale the experimental distribution $N$ as $N^{'} = N-M(\phi)$ (Fig. \ref{fig:nDTcut2}). 
We considered only tracks with $N^{'} \! \! <3 \sigma$ (one-sided cut), where $\sigma$ is the standard deviation of the Gaussian fit to $N^{'}$. We verified by visual inspection that events rejected by the latter cut are characterized by a large number of additional fired tubes in the neighbourhood of the correct ones.

\begin{figure}[b]
\begin{center}
\includegraphics[width=1.0\columnwidth]{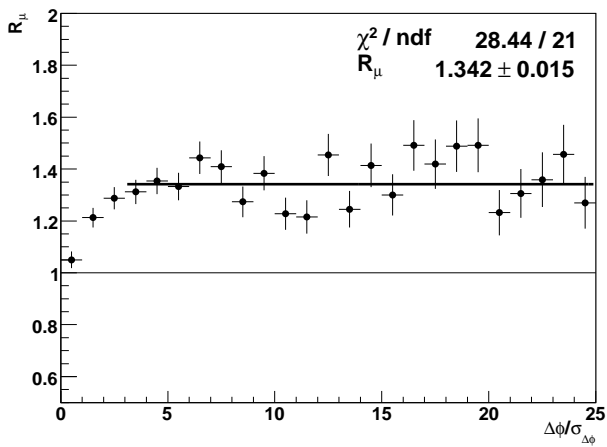}
\end{center}
\caption{Dependence of the measured charge ratio $R_\mu$ on the deflection angle expressed in units of experimental resolutions. A cut at $\Delta \phi / \sigma_{\Delta \phi}>$ 3 was applied in the data analysis. Note that the fitted value $R_\mu = 1.342 \pm 0.015$ was obtained with the bins indicated in the plot (the first 3 bins have not been used).}
\label{fig:cut2a}
\end{figure}

\begin{figure*}[t]
   \centerline{\subfloat[Before deflection cut]{\includegraphics[width=1.0\columnwidth]{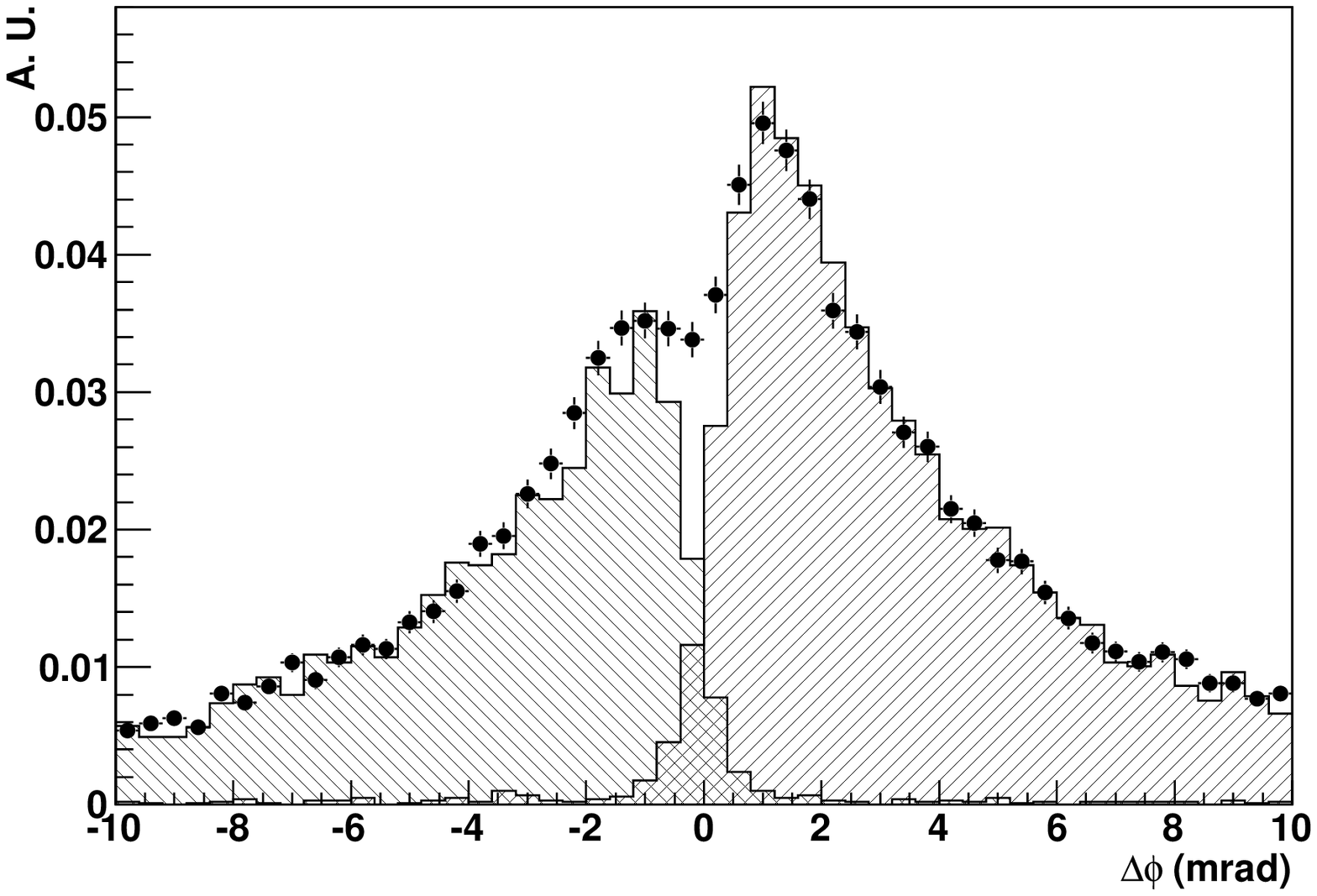} \label{fig:cut2b1}}
              \hfil
              \subfloat[After deflection cut]{\includegraphics[width=1.0\columnwidth]{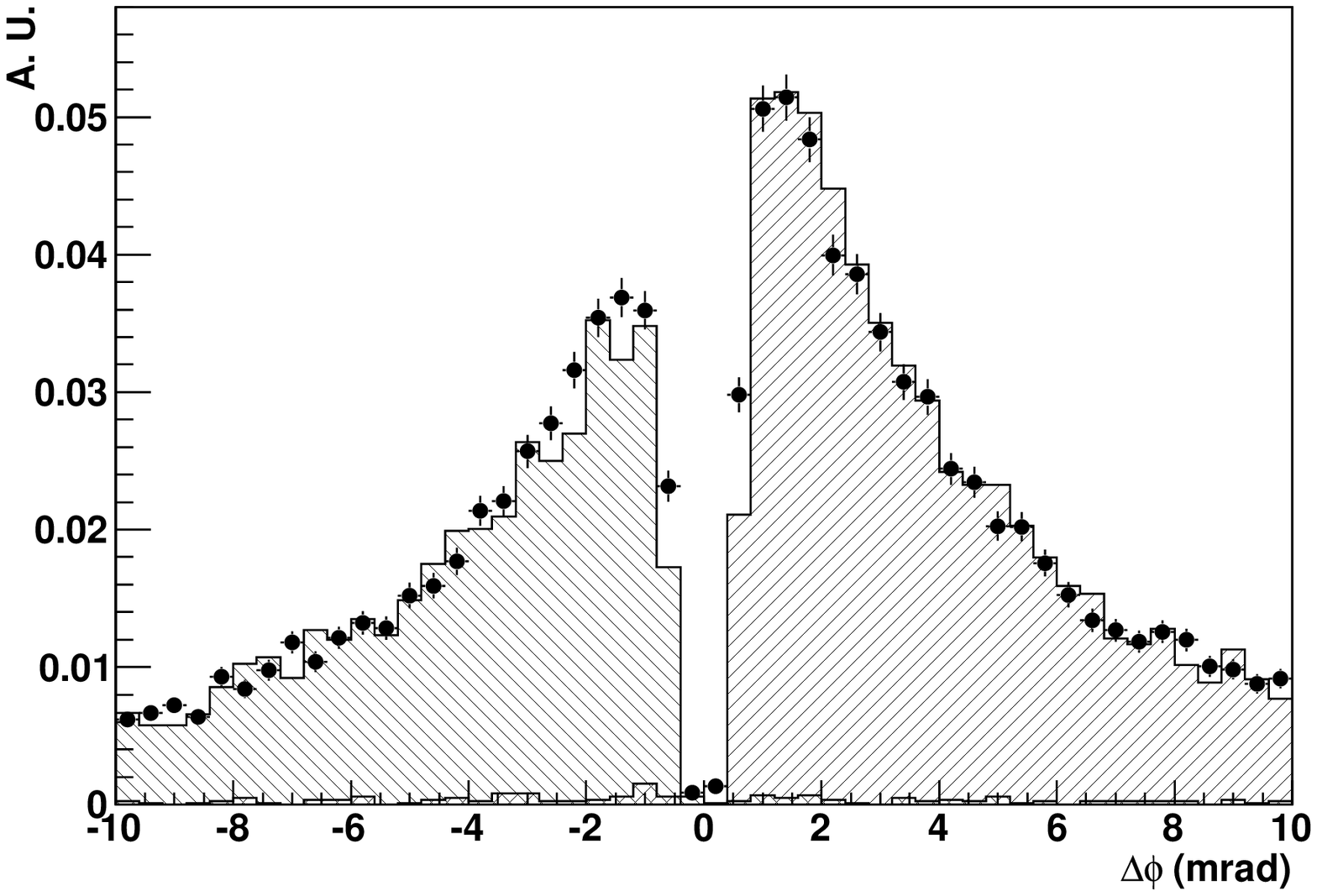} \label{fig:cut2b2}}
             }
   \caption{Effect of the deflection cut on $\Delta \phi$ distributions reconstructed exclusively from doublets (shown in the range -10 $\div$ 10 mrad). On the left is shown the distribution before the cut, where the two peaks corresponding to $\mu^{+}$ and $\mu^{-}$ are already clearly visible. Black points correspond to experimental data, hatched histograms to Monte Carlo simulations, split in the two components $q_{true} >$ 0 and $q_{true} <$ 0. The same distributions are shown on the right after the application of the deflection cut $\Delta \phi / \sigma_{\Delta \phi} >$ 3. The overlapping region of the two hatched histograms corresponds to the charge-misidentified tracks.}
   \label{fig:cut2b}
\end{figure*}

\begin{table*}[t]
\centering
\begin{tabular}{@{}|l|ccc|ccc|}
\hline
                &           & Data  &         &         & MC1     &         \\
                & evt/day & f$_{1}$ & f$_{2}$ & evt/day & f$_{1}$ & f$_{2}$ \\ 
\hline
Acceptance      & 992 & 100.0\%  & -      & 1222 & 100.0\% & - \\
Clean PT        & 515 &  51.9\%  & -      &  959 &  78.5\% & - \\
Deflection      & 391 &  39.4\%  & 76.0\% &  708 &  58.0\% & 73.8\% \\
\hline
Single $\mu$    & 379 &  38.2\%  & 96.9\% &  673 &  55.1\% & 95.1\% \\
Multiple $\mu$  &  12 &   1.2\%  &  3.1\% &   35 &   2.9\% &  4.9\% \\
\hline
\end{tabular}
\caption{Progressive reduction of the number of events per day after each selection cut, for data (left) and for MC1 (right). The effect of data reduction is also shown by reporting the fraction of events referred to the original sample (f$_{1}$) and to the previous cut (f$_{2}$). The total number of experimental events surviving the cuts is 44370.}
\label{tab:tab1}
\end{table*}

A further cut was applied on the $\Delta \phi$ angle ({\it deflection cut}). Events having a $\Delta \phi$ smaller or compatible with the experimental resolution were rejected. On the basis of the plot shown in Fig. \ref{fig:cut2a} where as expected for small deflection values $R_\mu \rightarrow 1$, only events with $\Delta \phi / \sigma_{\Delta \phi} >$ 3 were selected. The effect of this cut is visible in Fig. \ref{fig:cut2b} in which the $\Delta \phi$ distribution is shown before and after its application. In these plots, experimental data (black points) are plotted with the corresponding Monte Carlo distributions split in the two regions corresponding to positive particles ($q_{true} >$ 0) and negative particles ($q_{true} <$ 0).
The charge-misidentification $\eta$ corresponds to the overlapping region of the two distributions. Averaged over all the event samples $\eta$ is reduced from 0.080$\pm$0.002 to 0.030$\pm$0.001 by this cut.
The source of events with large $\Delta \phi$ angles and reconstructed with wrong charge-sign was investigated. A visual scan of Monte Carlo events confirms the hypothesis that they are due to secondary particles in the neighbourhood of the true muon track: if the two tracks are very close, it may happen that the track reconstructed with the best $\chi^2$ is the wrong one.
A further selection $\Delta \phi <$ 100 mrad was used to reject these fake tracks, with a small impact on the statistics. This last selection affects the sample with $p_\mu \aprle$ 5 GeV/c.

The muon charge ratio was computed separately for single muon events (i.e. event multiplicity $n_\mu$ = 1) and multiple muon events ($n_\mu >$ 1). Single muon events are selected by requiring single tracks in each projected view merged in the three dimensional space. Multiple muon events are selected by requiring a muon multiplicity $\ge\,$2 in both views, with tracks identified and unambiguously  merged in 3D space.

Tab. \ref{tab:tab1} lists the number of events remaining at each stage of the selection process. Note that data and MC1 event rates are absolute (given in day$^{-1}$) and not normalized one to the other. Also note that the {\it clean PT} cut has a stronger impact on data reduction and that the effect on the experimental data is different from that of Monte Carlo. This was expected since the percentage of events with PT digits not related to the muon track is intrinsically larger in the experimental data. The {\it clean PT} cut was tuned in order to be left with a clean data sample at the expense of a considerable loss of statistics.

\subsection{Alignment of the PT system}
\label{subsec:alignment}
The measurement of the muon charge is strongly affected by the alignment precision of the PT system. Misalignment effects have ``global'' or ``local'' contributions. To correct for global effects, which are the dominant ones, each station is treated as an independent rigid body and relative rotations and translations of one station with respect to the others are searched for. The local misalignment contribution takes into account possible distortions or bendings within each station.

A first alignment campaign was carried out with a theodolite to measure the position of the PT walls in the OPERA coordinate system. Recently a more refined alignment using cosmic ray muons was performed. The alignment procedure was carried out in two steps a) PT stations forming a doublet were aligned with the whole data sample, since 
the space in between has no magnetic field and tracks do not suffer any deflection; b) each doublet (pair) treated as a unit, separated by the iron magnet arm, was aligned using special runs with the magnetic field switched off.
This procedure allowed aligning two PT stations within a doublet with a spatial accuracy of $\sim$0.1 mm and an angular accuracy of $\sim$0.1 mrad, and to align two doublets with an angular accuracy of 0.2 mrad.

Local effects, such as bendings or distortions, contribute at the second order level and due to the present limited statistics have not been corrected for. However in Sec. \ref{subsec:syst} we provide an estimate of the systematic uncertainty on $R_\mu$ introduced by these effects.

\section{Underground muon charge ratio}
\label{sec:cr}

\subsection{Computation of $R_\mu$}
\label{subsec:cr}
$R_\mu$ was computed separately for single and multiple muon events.
Tab. \ref{tab:tab2} refers to single muon events where the number of positive and negative muons, their ratio, the charge-misidentification $\eta$ and the unfolded charge ratio are reported.
The $\eta$ value, defined as the fraction of tracks reconstructed with wrong charge sign, was extracted from the Monte Carlo simulation. Once $\eta$ is known, the unfolded charge ratio is obtained
according to the formula (see Appendix \ref{app:unfolding}):
\begin{equation}
R^{unf}_{\mu} = \frac{(1-\eta)R^{meas}_{\mu}-\eta}{-\eta R^{meas}_{\mu}+(1-\eta)}
\label{eq:unfold}
\end{equation}
The single muon sample was subdivided into three classes: tracks reconstructed exclusively as doublets, tracks reconstructed exclusively as singlets and as mixed. We verified that the fraction of these classes for experimental data and for Monte Carlo simulation are compatible: 54.8\% (doublets), 9.0\% (singlets) and 36.2\% (mixed) for real data to be compared to 52.5\%, 10.0\% and 37.5\% for Monte Carlo simulation (the errors are $\aprle$ 0.5\%). The final charge ratio value for single muon events, integrated over all the classes, is:
\begin{equation}
R^{unf}_{\mu}(n_\mu=1) = 1.377 \pm 0.014
\end{equation}
The same procedure was applied to multiple muon events. We selected events with $n_\mu > 1$ and reconstructed the charge of muons crossing the spectrometer section.
Events were classified in this category provided that more than one muon was reconstructed in the detector even though only one charge was measured.
In other words, the muon multiplicity is used to ``tag'' events generated by heavier and more energetic primaries.
The possibility to compute the muon charge ratio within the same event is presently excluded by the lack of statistics of high multiplicity muon bundles.

The charge ratio is $R^{meas}_{\mu}(n_\mu > 1)$ = 919/753 = 1.22 $\pm$ 0.06 and the corresponding unfolded value, obtained from Eq. \ref{eq:unfold}
\begin{equation}
R^{unf}_{\mu}(n_\mu > 1) = 1.23 \pm 0.06.
\end{equation}
This value is $2.4 \sigma$ away from the value for single muon events, consistent with the hypothesis of dilution of $R_\mu$ due to the neutron enhancement in the primary nuclei.

Tab. \ref{tab:tab3} gives information obtained with MC2 on some variables of single muon events and muon bundles in the OPERA detector. In particular, are given the average primary mass number $\langle A \rangle$, the average primary energy/nucleon ${\langle E/A \rangle}$, the fraction of Hydrogen nuclei over the total (H fraction), the
ratio of protons over neutrons in the primary radiation $N_p/N_n$ and finally the measured muon charge ratio $R^{unf}_\mu$.

\begin{table*}[t]
\centering
\begin{tabular}{@{}lccccc}
\hline
  & $N_{\mu^{+}}$  & $N_{\mu^{-}}$ & $R^{meas}_\mu$ & $\eta$ & $R^{unf}_\mu$ \\
\hline
Doublets      & 13595 & 9993 & 1.360 $\pm$ 0.018 & 0.0165 $\pm$ 0.0012 & 1.375 $\pm$ 0.019  \\
Mixed         &  8951 & 6603 & 1.355 $\pm$ 0.022 & 0.0403 $\pm$ 0.0022 & 1.393 $\pm$ 0.025  \\
Singlets      &  2181 & 1704 & 1.28  $\pm$ 0.064 & 0.064  $\pm$ 0.005  & 1.33  $\pm$ 0.05  \\
\hline
\end{tabular}
\caption{Final statistics for the underground muon charge ratio. Results are given separately for the three classes of events defined in the text. Errors are statistical only.}
\label{tab:tab2}
\end{table*}

\begin{table*}[t]
\centering
\begin{tabular}{@{}lccccc}
\hline
$n_\mu$ & $\langle A \rangle$ & ${\langle E/A \rangle}_{primary}$  & H fraction       & $N_p/N_n$      & $R^{unf}_\mu$   \\
\hline
  =1    & 3.35$\pm$0.09       & (19.4$\pm$0.1) TeV                 & 0.667$\pm$0.007  & 4.99$\pm$0.05  & 1.377$\pm$0.014 \\
$>$1    & 8.5$\pm$0.3         & (77$\pm$1) TeV                     & 0.352$\pm$0.012  & 2.09$\pm$0.07  & 1.23$\pm$0.06   \\
\hline
\end{tabular}
\caption{Primary cosmic ray information for single and multiple muon events (see text). Reported numbers were obtained with MC2 and with the composition model fitted in \cite{hoerandel}.
Only statistical errors are quoted. Systematic uncertainties related to the composition model dominate and can be inferred from the cited reference ($\delta \langle A \rangle \simeq 1$). In the last column the measured (and unfolded) charge ratios are given.
}
\label{tab:tab3}
\end{table*}

\begin{table*}[t]
\centering
\begin{tabular}{@{}cccccccc}
\hline
  Bin  &  $\mathcal{E}_\mu \cos \theta^{*}$ range & $\langle \mathcal{E}_\mu \cos \theta^{*} \rangle$ & $N_{\mu^{+}}$  &  $N_{\mu^{+}}$  &  $R^{unf}_\mu$  &  $\delta R^{unf}_\mu$(stat)  &    $\delta R^{unf}_\mu$(sys) \\
       & (GeV) & (GeV) & & & & & (\%) \\
      
\hline
 1	&  891 -- 1259  & 1197 &   899	&   678	& 1.353	& 0.074 & 0.4 \\
 2	& 1259 -- 1778	& 1527 & 14125	& 10571	& 1.373	& 0.019 & 0.5 \\
 3	& 1778 -- 2512	& 2071 & 10345	&  7613	& 1.420	& 0.025 & 1.3 \\
 4	& 2512 -- 3548	& 2852 &  3232	&  2444	& 1.409	& 0.047 & 4.3 \\
 5	& 3548 -- 7079	& 4329 &   643	&   548	& 1.192	& 0.079 & 3.1 \\
\hline
\end{tabular}
\caption{Main information for the five bins in $\mathcal{E}_\mu \cos \theta^{*}$. From left to right: the energy range and average value, the number of muons reconstructed with positive and negative charges, the unfolded
charge ratio, the statistical and systematic errors.}
\label{tab:tab4}
\end{table*}

\subsection{Systematic uncertainty on $R_\mu$}
\label{subsec:syst}
The main sources of systematic uncertainties in the determination of $R^{unf}_\mu$ are related to the alignment accuracy of the PT system and to the determination of the $\eta$ value.

\begin{figure*}[t]
   \centerline{\subfloat[Spectrometer of SM1]{\includegraphics[width=1.0\columnwidth]{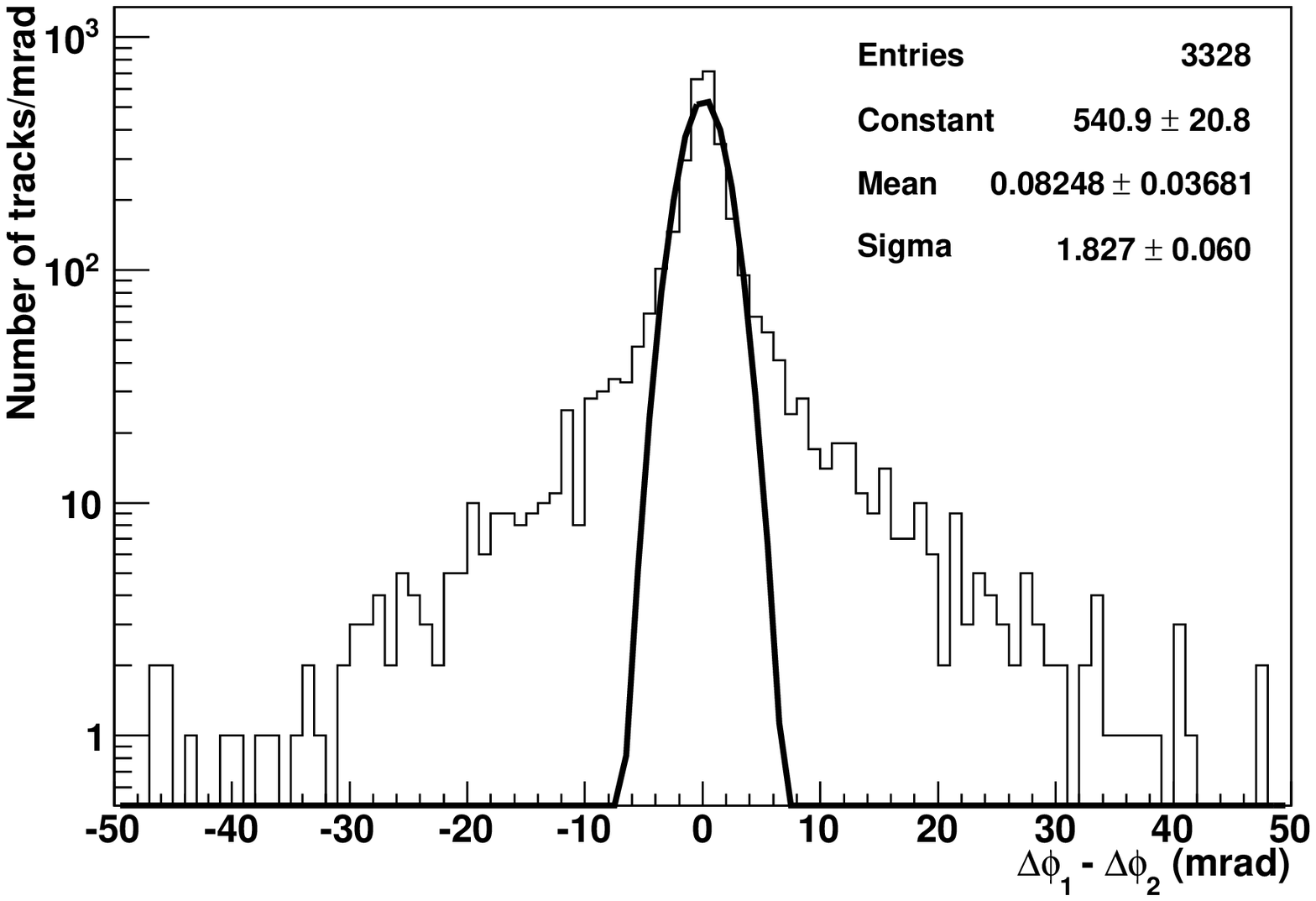} \label{fig:twoarm1}}
               \hfil
               \subfloat[Spectrometer of SM2]{\includegraphics[width=1.0\columnwidth]{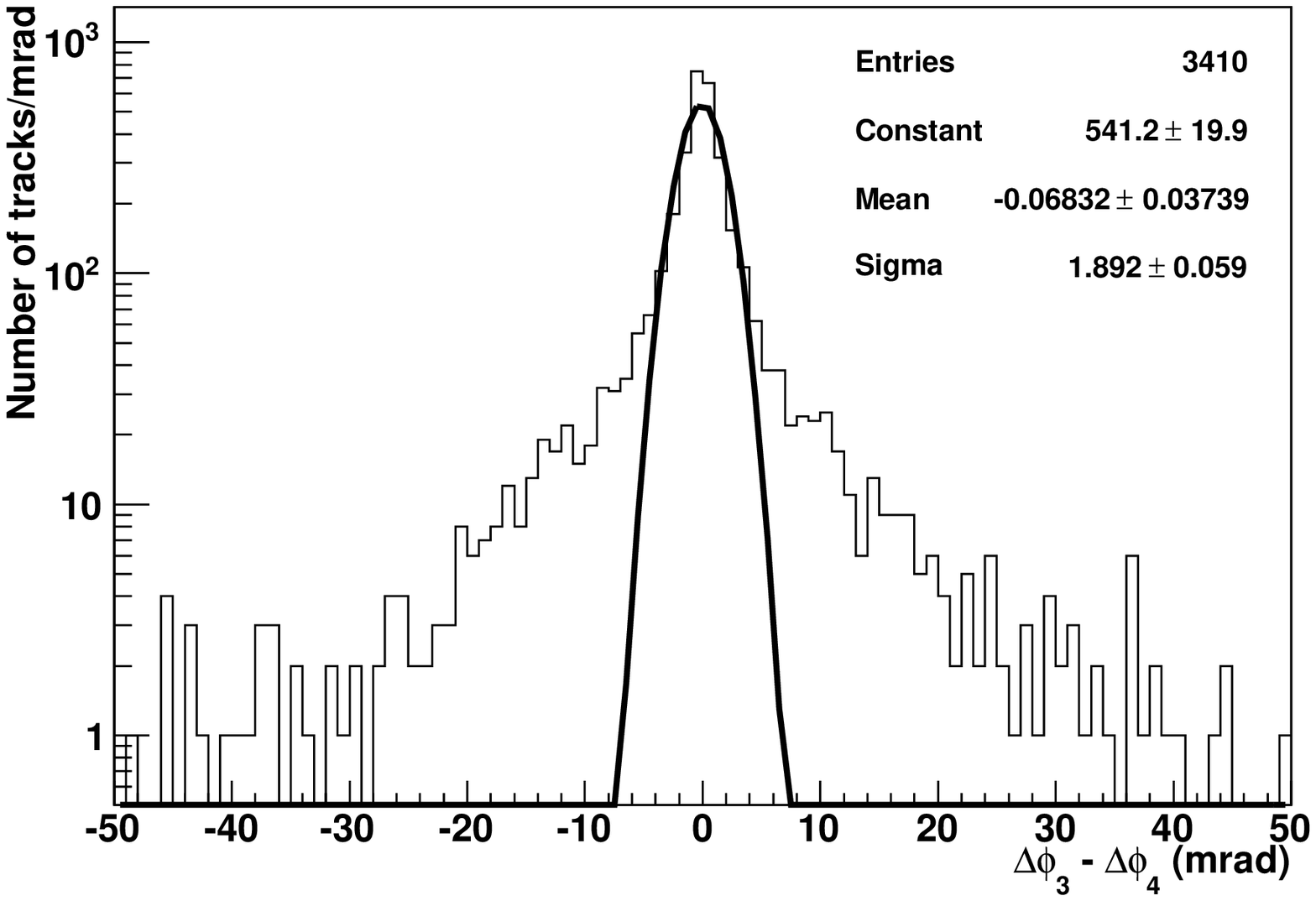} \label{fig:twoarm2}}
             }
   \caption{Two-arm test. Distributions of the difference of the deflection angles for tracks crossing both arms of one spectrometer: SM1 (left) and SM2 (right). In each plot we show the fit of the central part of the distributions to a Gaussian function.}
   \label{fig:twoarm}
\end{figure*}

The systematic uncertainty due to misalignment effects was evaluated in different ways. A given offset $\Delta \phi \rightarrow \Delta\phi + \delta\phi$ can be directly propagated in the algorithm which computes the charge ratio to evaluate $R_\mu \rightarrow R_\mu + \delta R_\mu$.
The $\delta \phi$ = 0.2 mrad uncertainty on the alignment accuracy obtained with magnets off (Sec. \ref{subsec:alignment}) corresponds to $\delta R_\mu$ = 0.03. However a more powerful procedure was used to better estimate this systematics. We considered all muon tracks crossing both arms of each spectrometer, thus providing two independent deflection values $\Delta \phi$ per spectrometer for the same muon track. With perfect alignment and neglecting the energy loss the difference $\delta \Delta \phi = \Delta \phi_{arm_1} - \Delta \phi_{arm_2}$
should be peaked at zero. The two distributions, one for each spectrometer, are shown in Fig. \ref{fig:twoarm} together with a Gaussian fit to the central part of the distributions, where the effects of muon energy loss in the magnet iron are negligible. The two peaks are at 0.08 mrad and -0.07 mrad respectively, $\sim$2 standard deviations away from zero. A misalignment of 0.08 mrad produces an error on the charge ratio $\delta R_\mu \simeq$ 0.015. We quote this number as the limiting alignment accuracy of each doublet with respect to the other. This number is conservative since it assumes that all four arms are affected independently from the same uncertainty. In reality only the the outer two doublets of each magnet contribute to this error, since a given offset in the central doublet	cancels the systematic uncertainty for $\Delta \phi_{arm_1}$ and $\Delta \phi_{arm_2}$.

A further test which also incorporates local effects consists in comparing the values $R^{i}_\mu$ ($i$=1,...,4) in each magnet arm. 
The average difference from the mean value $\sum_i | R^{i}_\mu - \bar{R}_\mu| / 4$ = 0.017 is within the statistical accuracy of each
$\delta R^{i}_\mu$ = 0.03.

Another consistency check exploited a small data sample ($\sim$9 days of livetime) obtained after inverting the polarity of the magnetic field. Running with inverted magnetic polarity could in principle cancel the systematic error related to misalignment effects. The result is $R^{inverted}_\mu$ = 1.36 $\pm$ 0.04, corresponding to the unfolded value $R^{inverted}_\mu$ = 1.39 $\pm$ 0.04. Even if the statistical error is larger than the systematic error quoted above, the result is in good agreement with the value obtained with normal polarity.

The charge-misidentification $\eta$ was previously estimated using Monte Carlo simulations. As already discussed the value is larger than what is expected from multiple scattering alone.
The difference is ascribed to the inclusion of spurious effects, such as the production of secondary particles near the muon trajectory, timing errors, and other second order effects not reproducible with the Monte Carlo program. Therefore we expect that the systematic uncertainty on $\eta$ is one-sided, being $\eta_{real} \ge \eta_{MC}$. To estimate this difference $\eta$ was evaluated using experimental data for a subsample of events. We considered all muon tracks crossing both arms of each spectrometer, which provide two independent deflections $\Delta \phi$ of the same muon track. In this case, the probability that the two deflection angles have opposite sign is $p = 2\eta(1-\eta)$ and therefore $\eta = 1 - \sqrt{1-2p}$. This formula neglects the correlation between the two $\Delta \phi$ angles, since they are built using a common track (the one in between the two arms). The correct $\eta(p)$ relation was derived using a Monte Carlo simulation
applied to the experimental and simulated data. It was found for the case of doublets $\eta_{data} = 0.018\pm0.002$ and	 $\eta_{MC} = 0.012\pm0.002$. Considering doublets and mixed configuration together, we found $\eta_{data} = 0.026\pm0.002$ and $\eta_{MC} = 0.019\pm0.002$. The difference $\delta \eta = 0.007$ was used as an estimate of the systematic uncertainty on $\eta$ which corresponds to $\delta R_\mu$ = 0.007.

The final systematic error is taken as the quadratic sum of its contributions and it is assumed be the same for single and for multiple muon events:
\begin{equation}
\delta R^{unf}_{\mu}(\textrm{syst.}) = ^{+ 0.017} _{-0.015}
\end{equation}

\section{$R_\mu$ as a function of $p_\mu$ and $\mathcal{E}_\mu \cos \theta^{*}$}
\label{sec:final}
\begin{figure}[b]
\begin{center}
\includegraphics[width=1.0\columnwidth]{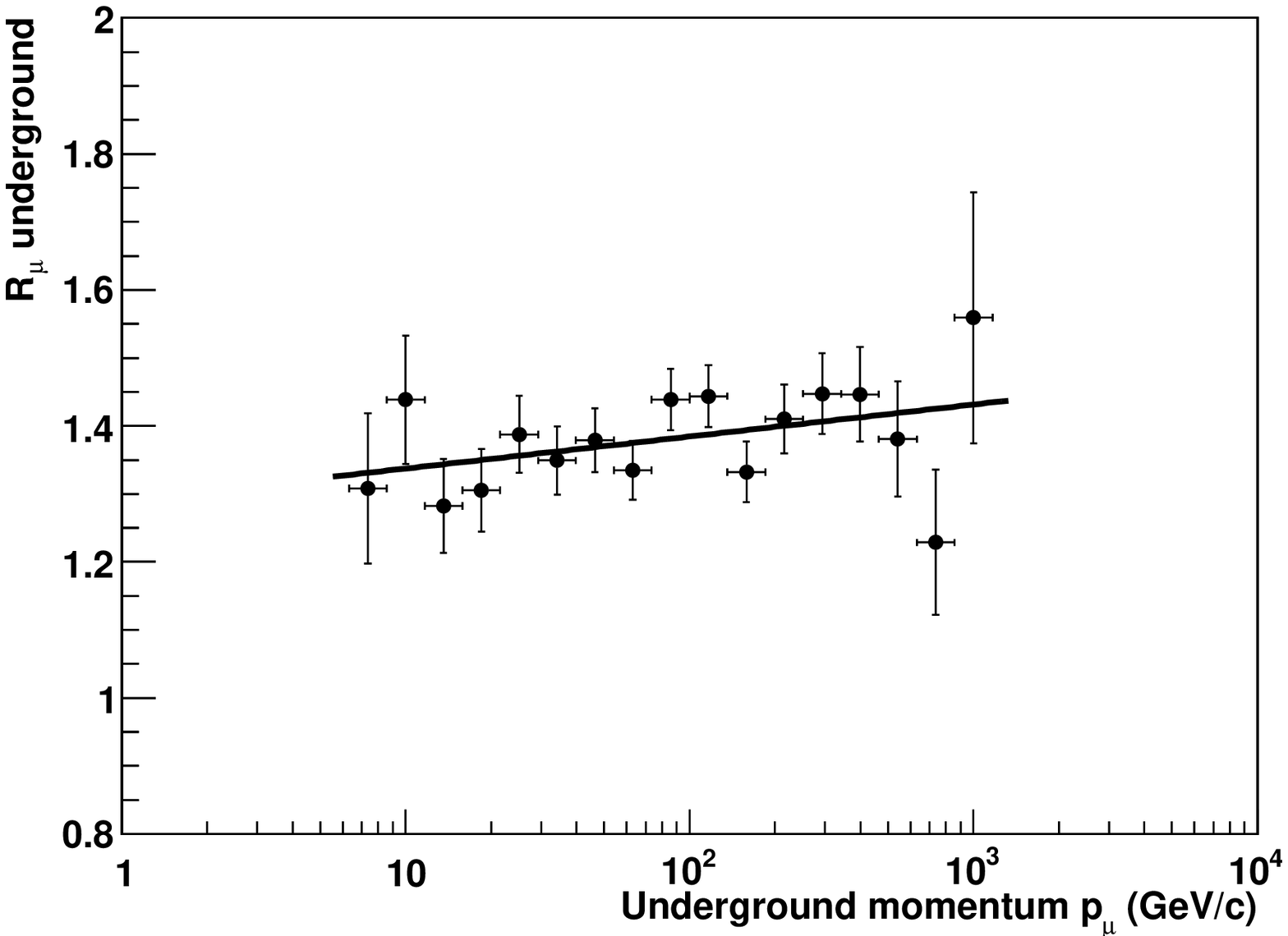}
\end{center}
\caption{Measured charge ratio of underground muons as a function of the reconstructed muon momentum. Data points below $\sim$5 GeV/c and above $\sim$1000 GeV/c are suppressed by the cut on $\Delta \phi$.
A fit of the form $R_\mu (p_\mu) = a_0 + a_1 \mathrm{log_{10}}[p_\mu/\mathrm{(GeV/c)}]$ is superimposed to the data.}
\label{fig:edown}
\end{figure}

The underground muon momentum $p_\mu$ was computed using Eq. \ref{eq:momentum}.
The muon charge ratio as a function of $p_\mu$ is shown in Fig. \ref{fig:edown}, where the widths of the horizontal error bars correspond approximately to the (average) muon momentum resolution.
A linear fit 
\begin{equation}
R_\mu (p_\mu) = a_0 + a_1 \mathrm{log_{10}}[p_\mu/\mathrm{(GeV/c)}]
\end{equation}
gives $a_0 = 1.29 \pm 0.06$ and $a_1 = 0.05 \pm 0.03$ with $\chi^2/dof = 13.7/15$. The data are also compatible with the hypothesis of a 
constant charge ratio, since the fit to a constant yields $a_0 = 1.379 \pm 0.015$ with $\chi^2/dof = 16.2/16$ and therefore $\Delta \chi^2/dof = 2.47/1$ (corresponding to $\sim$1.6 sigma).

\begin{figure*}[t]
\begin{center}
\includegraphics[width=2.0\columnwidth]{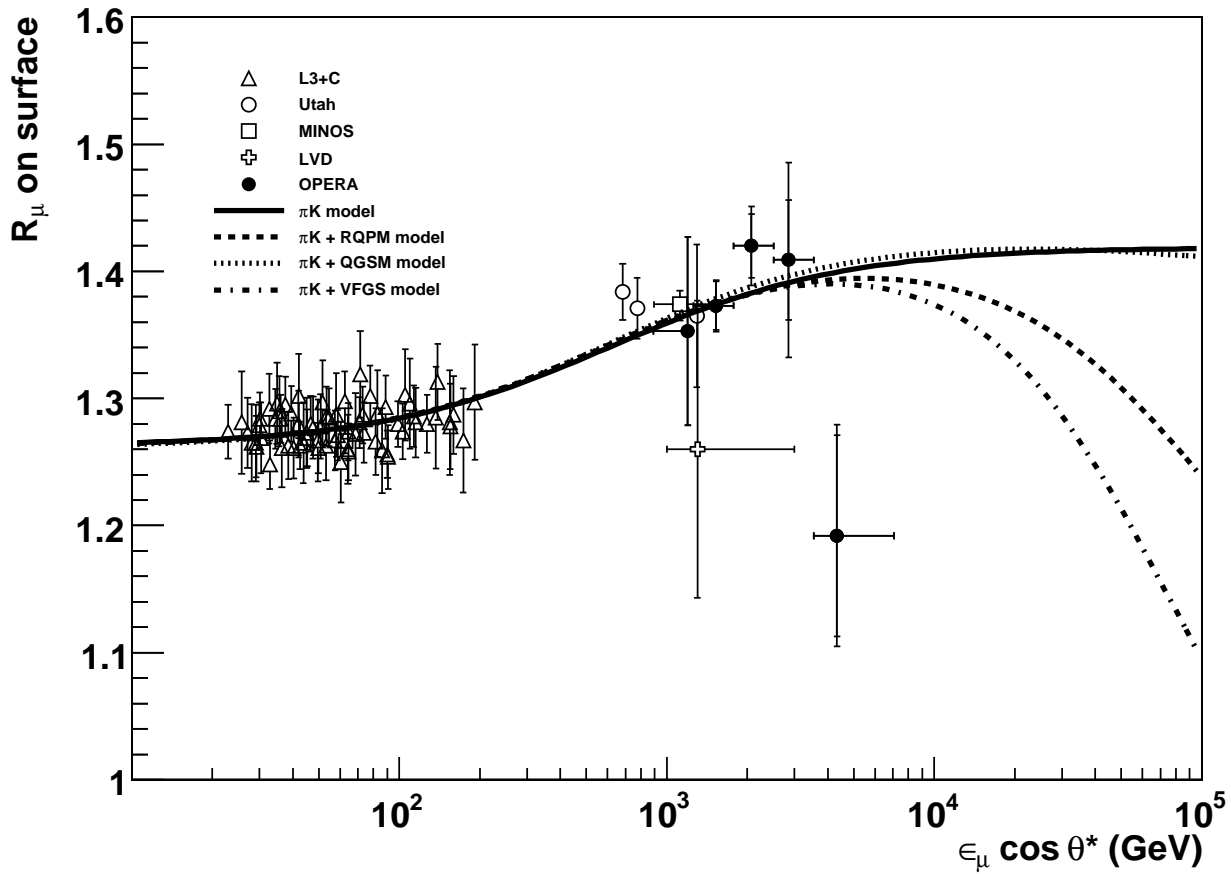}
\end{center}
\caption{$R_\mu$ values measured by OPERA in bins of $\mathcal{E}_\mu \cos \theta^{*}$ (black points). Also plotted are the data in the low energy region from MINOS-ND \cite{minos-nd} and L3+C \cite{l3c} and in the high energy region from Utah \cite{utah}, MINOS \cite{minos} and LVD \cite{lvd} experiments. The result of the fit of OPERA and L3+C data to Eq. \ref{eq:fit} is shown by the continuous line. The dashed, dotted and dash-dot lines are, respectively, the fit results with the inclusion of the RQPM, QGSM \cite{bugaev} and VFGS \cite{volkova} models for prompt muon production in the atmosphere.}
\label{fig:emucos}
\end{figure*}

The muon energy at the surface ($\mathcal{E}_\mu$) is directly related to the underground residual energy ($E_\mu \simeq p_\mu$) and to the rock amount crossed by the muon to reach the detector level. In fact, the energy loss of high energy muons in the rock is usually expressed as
\begin{equation}
-\frac{dE}{dh} = \alpha(E) + \beta(E) E
\label{eq:dedx}
\end{equation}
where $h$ is the rock depth while the two energy-dependent parameters $\alpha$ and $\beta$ are the contributions of the ionization energy loss and the radiative processes, respectively. Eq. \ref{eq:dedx} can be integrated to obtain the approximate formula
\begin{equation}
\mathcal{E}_\mu = (E_\mu + \alpha/\beta)e^{\beta h} - \alpha/\beta
\label{eq:proj}
\end{equation}
which connects the surface and underground muon energies. However, Eq. \ref{eq:proj} is valid only on average. The ``resolution'' $d\mathcal{E}_\mu = \mathcal{E}^{rec}_\mu - \mathcal{E}^{true}_\mu$ is dominated by the statistical fluctuations due to the discrete processes described by the term $\beta$ \cite{lipari2}. We evaluated $\mathcal{E}_\mu$ with a full Monte Carlo simulation to build the table $\mathcal{E}_\mu$ = $f(h,p_\mu)$. For this purpose the code MC2 was used since it contains a detailed description of the muon flux at the surface and the muon transport in the Gran Sasso rock. The $(h,p_\mu)$ plane was divided into 10$\times$10 equally-spaced bins and in each bin the average $\langle \mathcal{E}_\mu \rangle$ value was computed. The binning was chosen coarse enough to have a large statistical sample in each bin without affecting the resolution $d\mathcal{E}_\mu$, which is of the order of 0.15 in the logarithmic scale.
The surface muon charge ratio was computed as a function of the variable $\langle \mathcal{E}_\mu \rangle \cos \theta^{*}$ binned according to the resolution.
Finally, the experimental values were corrected in each bin for the corresponding charge mis-identification and shown in Fig. \ref{fig:emucos} with black points (only single muon events were considered).
The present statistics does not allow to draw conclusions about the highest energy data point shown in the figure.

Tab. \ref{tab:tab4} gives some information for each of the five bins considered: the energy range and average value, the statistical sample, the unfolded charge ratio, the statistical and systematic errors.
The latter were evaluated computing in each bin the two contributions discussed in Sec. \ref{subsec:syst}.

In Fig. \ref{fig:emucos} are shown for comparison the data from other experiments for which we could recover information on the $\mathcal{E}_\mu \cos \theta^{*}$ variable. For the low energy region we took data from Ref. \cite{minos-nd} and Ref. \cite{l3c} (we choose data points with uncertainties $\delta R_{\mu}$$<$$0.05$) while in the high energy region the data are from Ref. \cite{utah} and Ref. \cite{minos}. For the latter, since the angular information were not provided in the paper, we plotted the $R_\mu$ integrated value in correspondence of the $\mathcal{E}_\mu \cos \theta^{*}$ value given in Ref. \cite{icrc07-goodman}. We also report a recent result from Ref. \cite{lvd} where the vertical muon charge ratio is given in the range 1-3 TeV (average value 1.3 TeV).

Finally, we fit our data to Eq. \ref{eq:2}, using a procedure similar to what is described in Ref. \cite{minos}. We rewrite Eq. \ref{eq:2} in the form:

\begin{equation}
\phi_{\mu^{\pm}} \! \propto \! \frac{a_\pi f_{\pi^{\pm}}}{1+b_\pi \mathcal{E}_\mu \cos \theta^{*} /\epsilon_{\pi}} + R_{K\pi} \frac{a_K f_{K^{\pm}}}{1+b_K \mathcal{E}_\mu \cos \theta^{*} /\epsilon_{K}}
\label{eq:fit}
\end{equation}
where $R_{K\pi} = Z_{NK}/Z_{N\pi}$ and $f_{\pi^{+}} = 1 - f_{\pi^{-}} = Z_{N\pi^{+}}/Z_{N\pi}$ (and similarly for kaons).
$f_{\pi^{+}}$ and $f_{K^{+}}$ were left free to vary while we fixed the kinematical parameters $a_{\pi}$ = 0.674, $a_{K}$ = 0.246, $b_{\pi}$ = 1.061, $b_{K}$ = 1.126 and the fraction of kaons over pions in the atmosphere $R_{K\pi} = 0.149$ \cite{gaisser}. The fit of $R_\mu = \phi_{\mu^{+}}/\phi_{\mu^{-}}$ takes into account data from \cite{minos-nd} and \cite{l3c} for the low energy region and data from this work at higher energies. The fit yields the values $f_{\pi^{+}}$ = 0.5514$\pm$0.0014 and $f_{K^{+}}$ = 0.680$\pm$0.015 which correspond to a ratio $R_\pi = Z_{N\pi^{+}}/Z_{N\pi^{-}}$ = 1.229$\pm$0.001 and $R_K = Z_{NK^{+}}/Z_{NK^{-}}$ = 2.12$\pm$0.03 for pions and kaons respectively. The result of the fit is shown in Fig. \ref{fig:emucos} as a continuous line.

The contribution of the prompt muon component to $R_\mu$ was evaluated for three different charm production models: the phenomenological non-perturbative models \linebreak RQPM and QGSM \cite{bugaev} and the semi-empirical model from Volkova {\em et al.} \cite{volkova}. In \cite{bugaev} the prompt muon flux and charge ratios are parametrized as a function of the muon energy. The results of the fit extended to include the prompt contribution as predicted by these models, are shown in Fig. \ref{fig:emucos}. The pion and kaon charge ratios obtained from the fit are unchanged within the statistical errors.

\section{Conclusions}
\label{sec:conclusion}
The atmospheric muon charge ratio $R_{\mu} = N_{\mu^{+}}/N_{\mu^{-}}$ was measured using the spectrometers of the OPERA underground detector. We analyzed four months of data taken during the 2008 CNGS neutrino run.  
For single muons the $R_\mu$ value integrated over the underground muon spectrum is
\begin{equation}
R^{unf}_{\mu}(n_\mu=1) = 1.377 \pm 0.014 \, (\textrm{stat.}) ^{+ 0.017} _{-0.015} \, (\textrm{syst.}) \nonumber
\label{eq:conclusion}
\end{equation}
to be compared to $R^{unf}_{\mu}(n_\mu > 1) = 1.23 \pm 0.06$ for muon bundles. This difference of about $\sim$2.4$\sigma$ supports the hypothesis of the decrease of the muon charge ratio with increasing primary mass. This is the first indication of such an effect which provides a further handle for the correct understanding and modelling of the secondary production in the atmosphere.

The underground muon charge ratio is consistent with past measurements in a similar energy region. Data suggest a slight increase of $R_\mu$ with the underground muon momentum, although a fit to a constant charge ratio cannot be excluded. 

The dependence of $R_\mu$ on the vertical surface energy $\mathcal{E}_\mu \cos \theta^{*}$
shows an increase in the region 1-3 TeV and it is compatible with a model which considers only the $\pi$ and $K$ contributions to the muon charge ratio. A fit of the low energy data and our data with a simplified description of the atmospheric muon flux provides a value of the pion and kaon charge ratios $R_\pi = Z_{N\pi^{+}}/Z_{N\pi^{-}}$ = 1.229$\pm$0.001 and $R_K = Z_{NK^{+}}/Z_{NK^{-}}$ = 2.12$\pm$0.03, respectively. The inclusion of the prompt muon component does not modify the fit results. It is however intriguing to observe that our measurement lies in the region where the charmed particle production may start to give an observable contribution to the muon charge ratio. A larger statistical sample or an experimental measurement with a new detector at very large depths could shed light on the region $\mathcal{E}_\mu \cos \theta^{*} \aprge$ 10 TeV. The data collected by OPERA at the end of its scientific program will allow to improve the measurement in this energy region.

\section*{Acknowledgments}
We thank INFN for the continuous support given to the experiment during the
construction, installation and commissioning phases through its LNGS laboratory. 
We warmly acknowledge funding from our national agencies: Fonds de la Recherche Scientifique - FNRS and
Institut Interuniversitaire des Sciences Nucleaires for Belgium, MoSES for Croatia, IN2P3-CNRS for
France, BMBF for Germany, INFN for Italy, the Japan Society for the Promotion of Science
(JSPS), the Ministry of Education, Culture, Sports, Science and Technology (MEXT) and the
Promotion and Mutual Aid Corporation for Private Schools of Japan for Japan, SNF and ETH Zurich
for Switzerland, the Russian Foundation for Basic Research (grants 08-02-91005-CERN\_a, SS 959.2008.2) for Russia,
the Korea Research Foundation Grant (KRF-2008-313-C00201) for Korea. We also thank
the INFN for providing fellowships and grants to non Italian researchers.
We are indebted to our technical collaborators for the excellent quality of their work over
many years of design, prototyping and construction of the detector and of its facilities.
Finally, we warmly acknowledge S.~Cecchini for fruitful discussions on cosmic ray physics.

\appendix

\section{Unfolded charge ratio}
\label{app:unfolding}
Let us call $m^{ij}$ the number of muons with charge $i$ reconstructed with charge $j$.
The total number of {\it true} positive and negative muons is therefore:
\begin{eqnarray}
M^{+} = m^{++} + m^{+-} \nonumber \\
M^{-} = m^{--} + m^{-+} \nonumber
\end{eqnarray}
On the other hand, the total number of {\it reconstructed} positive and negative muons is:
\begin{eqnarray}
\hat{M}^{+} = m^{++} + m^{-+} \nonumber \\
\hat{M}^{-} = m^{--} + m^{+-} \nonumber
\end{eqnarray}
Let us define the charge-misidentification $\eta$ as:
\begin{eqnarray}
\eta^{+-} = \frac{m^{+-}}{M^{+}} \nonumber \\
\eta^{-+} = \frac{m^{-+}}{M^{-}}
\end{eqnarray}
Using a matrix notation, we can express the relationship between $\mathbf{M}$ and 
$\mathbf{\hat{M}}$ as:
\begin{equation}
\mathbf{\hat{M}} = \mathbf{H} \mathbf{M}
\end{equation}
where
\begin{equation}
\mathbf{H} = 
\left\lgroup
\begin{array}{cc}
1-\eta^{+-} &   \eta^{-+} \\
  \eta^{+-} & 1-\eta^{-+}
\end{array}
\right\rgroup
\end{equation}
Inverting this relation, one has the number of ``true'' positive and negative muons:
\begin{equation}
\mathbf{M} = \mathbf{H}^{-1} \mathbf{\hat{M}}
\end{equation}
where
\begin{equation}
\mathbf{H}^{-1} = 
\frac{1}{1-\eta^{+-}-\eta^{-+}}
\left\lgroup
\begin{array}{cc}
1-\eta^{-+} &  -\eta^{-+} \\
 -\eta^{+-} & 1-\eta^{+-}
\end{array}
\right\rgroup
\end{equation}
The two $\eta$ values $\eta^{+-}$ and $\eta^{-+}$ are obtained from a Monte Carlo simulation. We found that,
within the statistical accuracy of the simulation,
$\eta^{+-}=\eta^{-+}=\eta$ as one would expect from a charge-symmetric detector.
This simplifies the expressions which, in terms of the ratio $R$, becomes
\begin{eqnarray}
R & = & \frac{M^{+}}{M^{-}} = \frac{(1-\eta)\hat{M}^{+}-\eta\hat{M}^{-}}{-\eta\hat{M}^{+}+(1-\eta)\hat{M}^{-}} = \nonumber \\
  & = & \frac{(1-\eta)\hat{R}-\eta}{-\eta\hat{R}+(1-\eta)}
\label{eq:unf}
\end{eqnarray}
where $\hat{R} = \hat{M}^{+}/\hat{M}^{-}$.

If $\hat{R}$ is computed with the same Monte Carlo events used to evaluate $\eta$, one would
obtain the same ``true'' $R$ value of the starting data sample. If $\hat{R}$ is computed with the
experimental reconstructed data, then $R$ is the unfolded experimental value in Eq. \ref{eq:unfold}.

The error $\delta R$ is obtained propagating the errors on $\hat{R}$ and $\eta$ over Eq. \ref{eq:unf}:
\begin{equation}
\delta R = \frac{\sqrt{(1-2\eta)^2 (\delta \hat{R})^2 + (\hat{R}^2-1)^2(\delta \eta)^2}}{[\eta \hat{R} - (1-\eta)]^2}
\end{equation}

It may be pointed out that we did not use any regularization scheme in the unfolding,
i.e. statistical fluctuations on $\hat{R}$ are not damped in Eq. \ref{eq:unf} in order to prevent unphysical spikes
in the unfolded $R$ value. This is acceptable in our case since the collected
statistics on $\hat{M}^+$ and $\hat{M}^-$ is large enough.

\end{document}